
\input phyzzx
\input tables
\nopagenumbers
\voffset = -0.4in
\footline={\ifnum\pageno=1 \nulline \else\newfootline \fi}
\def\nulline{{\hfill}}
\def\newfootline{\advance\pageno by -1\hss\tenrm\folio\hss}
\rightline {July 1993} \rightline {QMW--TH--93/21.}
\rightline {SUSX--TH--93/13.}
\title {Duality Symmetries
in Orbifold Models.}
\author{D. Bailin$^{a}$, \ A. Love$^{b}$,  \ W.A.
Sabra$^{b}$\ and \ S. Thomas$^{c}$}
\address {$^{a}$School of Mathematical and Physical
Sciences,\break
University of Sussex, \break Brighton U.K.}
\address {$^{b}$Department of Physics,\break
Royal Holloway and Bedford New College,\break
University of London,\break
Egham, Surrey, U.K.}
\address {$^{c}$
Department of Physics,\break
Queen Mary and Westfield College,\break
University of London,\break
Mile End Road, London,  U.K.}
\abstract { We derive the duality symmetries relevant to moduli
dependent gauge coupling constant threshold corrections, in Coxeter $ {\bf Z_N}
$
orbifolds. We consider those orbifolds for which the point group leaves fixed a
2-dimensional
sublattice $\Lambda_2$,  of the six dimensional torus lattice $\Lambda_6$,
where $\Lambda_6 $
cannot be decomposed as $\Lambda_2 \bigoplus{\Lambda_4}.$  }
\endpage
\REF\one{L. Dixon, J. A. Harvey, C. Vafa and E. Witten, Nucl. Phys.
B261 (1985) 678; B274 (1986) 285.}
\REF\two{ A. Font, L. E. Ibanez,
F. Quevedo and A. Sierra, Nucl. Phys. B331 (1991) 421.}
\REF\di{R. Dijkgraaf, E. Verlinde and H. Verlinde, Comm. Math.  Phys.115
	      (1988) 649.}
\REF\verlinde{ R. Dijkgraaf, E. Verlinde and H. Verlinde,
On Moduli Spaces of
Conformal Field
Theories with $c \geq 1$, Proceedings Copenhagen Conference,
Perspectives
in String Theory,
edited by P. Di Vecchia and J. L. Petersen,
World Scientific, Singapore, 1988.}
\REF\wil{ A. Shapere and F. Wilczek, Nucl.Phys. B320 (1989) 669.}
\REF\dine{ M. Dine, P. Huet and N. Seiberg,
Nucl. Phys. B 322 (1989) 301.}
\REF\la{ J. Lauer, J. Mas and H. P. Nilles, Nucl. Phys. B351 (1991) 353.}
\REF\jap{ K. Kikkawa and M. Yamasaki, Phys. Lett. B149, (1984) 357;
N. Sakai and I. Senda, Prog. Theor. Phys. 75 (1984)692}
\REF\giv {A. Giveon, E. Rabinovici and G. Veneziano,
Nucl. Phys. B322 (1989) 167.}
\REF\lust {W. Lerche, D. L\"ust and N. P. Warner,
Phys. Lett. B231 (1989) 417.}
\REF\erler {J. Erler and M. Spalinski, preprint, MPI-PH-92-61, TUM-TH-147-92,
(1992).}
\REF\threshold {V. S. Kaplunovsky, Nucl. Phys. B307 (1988) 145}
\REF\dixon {L. J.
Dixon, V. S. Kaplunovsky and J. Louis,  Nucl. Phys. B355 (1991) 649.}
\REF\ferrara{J. P. Derenddinger, S. Ferrara, C. Kounas and F. Zwirner, Nucl.
Phys. B372 (1992) 145, Phys. Lett. B271 (1991) 307.}
\REF\stie{ P. Mayr and S. Stieberger, MPI-Ph/93-07, TUM-TH-152/93 preprint,
to appear in Nucl. Phys.}
\REF\nar {K. S. Narain, Phys. Lett. B169 (1987) 41.}
\REF\sar{ K. S. Narain, M. H. Sarmadi  and E. Witten,   Nucl. Phys. B279 (1987)
369.}
\REF\spalinski {M. Spalinski, Nucl. Phys.
B377 (1992) 339.}
\REF\kat{Y. Katsuki, Y. Kawamura, T. Kobayashi, N. Ohtsubo, Y. Ono and K.
Tanioka, Nucl. Phys. B341 (1990) 611.}
\REF\us {D. Bailin, A. Love, W. A. Sabra and S. Thomas, SUSX--TH--93/14,
QMW--TH--93/22.}
\def\N{{\cal N}}
In the space of all known conformal field theories, orbifold models
represent good candidates for a phenomenologically promising string
compactification [\one, \two].
The marginal deformations of the underlying conformal
field theory of the orbifold are the moduli which parametrize, locally,
the string background vacuum [\di].
A peculiar feature in string compactifications, not shared with that of
conventional point-particles,
is the invariance of the spectrum under the action
of some discrete group acting on the moduli [\wil, \dine, \jap, \giv].
This  group,  the so-called target space duality which  generalizes
the well known
$R\rightarrow 1/R$ duality symmetry for circle compactification where $R$ is
the radius of the circle, can then be implemented to restrict the
moduli space to a fundamental domain.

The duality groups for toroidal and orbifold compactification in lower
dimensions
have been  considered in  [\di -\lust ].
For two-dimensional toroidal compactification [\verlinde] one finds two copies
of the modular group $PSL(2,Z)$ acting on the two complex moduli, $T $ and $U$
describing the target space. By comparison, for
 the two-dimensional ${\bf Z}_3$ orbifold, the $U$ modulus is frozen
(i.e.  its value is fixed), and the duality group $\Gamma_T$ associated with
the complex $T$ modulus is the modular group $PSL(2,Z)$.
 Based on these results, it was sometimes assumed in the literature that the
modular group is realized as a duality group for each complex modulus
associated with the three complex planes of the six-dimensional orbifold.
However a counter example has been found in ref. [\erler], in which the duality
group of the
Coxeter ${\bf Z}_7$ orbifold with $SU(7)$ lattice has been shown to be
an overall $PSL(2,Z)$ for the three complex moduli  $T_i$ rather than
$PSL(2,Z)^3$.

More recently in ref. [\stie], the moduli-dependent threshold corrections to
the gauge coupling constants for some orbifold models, arising from the twisted
sectors with one unrotated plane under the twist action   [\threshold, \dixon,
\ferrara], were found  not to be invariant
under the full modular group but rather under certain
congruence  subgroups of $PSL(2,Z)$.
These are the only sectors yielding moduli dependent
contributions  to the threshold corrections, and  are known as  $\N=2$ sectors
as they possess two space-time supersymmetries.
In ref. [\dixon] it was demonstrated that provided  the six-dimensional lattice
can be
 decomposed into a direct sum of a two-dimensional and a four-dimensional
sub-lattices, $\Lambda_6=\Lambda_2\bigoplus{\Lambda_4}$, with the unrotated
plane lying in $\Lambda_2$,
the threshold corrections are then invariant under the full modular group.
This is essentially consistent with the modular symmetry of those moduli
 associated with the plane lying in the two-dimensional lattice, $\Lambda_2$.
However many orbifold lattices do not admit the above decomposition.
Here we will investigate  the duality group
of the moduli of the invariant planes of these orbifolds, which is a symmetry
of those
 sectors of  Hilbert space contributing to  gauge coupling threshold
corrections.

This work is organized as follows.
First, we  briefly review  toroidal and orbifold compactification,
and the method of obtaining their duality symmetries.
Next we concentrate on the two-dimensional case and show
that the duality group is $PSL(2, Z)$ for all symmetric ${\bf Z}_N$ orbifolds.
This demonstrates that provided the lattice is a direct sum of three
two-dimensional sub-lattices, the duality group of
a six-dimensional orbifold is always a product of  the modular group
$PSL(2,Z)$, one for each complex modulus. Note that the $U$ moduli are only
present for  ${\bf Z}_2$ planes. The relevance
of  2-dimensional compactification to the study of  threshold corrections will
become
clear in what follows.

Finally,  motivated by the results of [\stie], we determine
the symmetry group  which leaves invariant the spectrum of the twisted sectors
with only two rotated planes, i.e. those that possess $\N = 2 $ supersymmetry.
This spectrum is only sensitive to the geometry of the unrotated complex plane
and independent of the moduli of the other two completely rotated complex
planes.
The  symmetry groups obtained are those relevant to threshold corrections of
the gauge coupling
constants  in  these models.
Only  the cases where the invariant planes do not lie entirely in a
two-dimensional sub-lattice of the 6-dimensional
 torus lattice,  a la Dixon et al [\dixon],
are considered.

We begin with some aspects of duality transformations of closed string
compactification on tori and orbifolds.  A $d$-dimensional torus is  defined as
a quotient of
${\bf R}^d$ with respect to a lattice
$ \Lambda$ defined by
$$\Lambda=\Big \{\sum_{i=1}^d a^ie_i,\qquad a^i\in Z\Big \}.\eqn\lat$$
In the absence of Wilson lines, the toroidal compactification [\nar, \sar] is
described by $d^2$ parameters given by an antisymmetric field $B_{ij}$, and  a
metric $G_{ij}$
defined as
$$G_{ij}=e_i.e_j \eqn\a$$

For the string coordinates compactified on the torus in the above
background, the left and right momenta are given by
$$P_L={p\over 2} +(G-B)w, \qquad P_R={p\over2}- (G+ B)w,\eqn\latt$$
where $w$ and $p$, the windings and the momenta respectively, are
$d$-dimensional integer valued vectors taking values on
the lattice $\Lambda$ and its dual $\Lambda^*$.
The zero modes of the compactified string coordinates which contains the
dependence on the
geometry of the background
has the contribution $H$ and $S$ to the scaling dimension and spin
of the vertex operators given by
$$\eqalign{H= &{1\over2} (P^t_L G^{-1} P_L + P^t_R G^{-1} P_R)=
{1\over4}p^tG^{-1}p-p^tG^{-1}Bw-w^tBG^{-1}Bw + w^tGw
,\cr
S= &{1\over2} (P^t_L G^{-1} P_L - P^t_R G^{-1} P_R)=p^tw,}\eqn\latt$$
It is very convenient to write $H$ and $S$ in the following quadratic forms
$$H={1\over2}  u^t \Xi u, \qquad S={1\over2} u^t \eta u.\eqn\scale$$
where
$$u = \pmatrix{w \cr p}, \quad  \eta = \pmatrix{0 & {\bf 1}_d \cr
	{\bf 1}_d & 0 },\quad
\Xi =\pmatrix{
	 2(G-B)G^{-1}(G+B) & BG^{-1}\cr
		     - G^{-1} B& {1 \over 2} G^{-1}}.\eqn\latti$$

Here the index $t$ denotes the transpose, $u$ is a $2d$ component integer
vector, $\Xi$ and $\eta$
are $2 d\times 2d $ dimensional matrices and ${\bf 1}_d$ denotes the
identity matrix in $d$ dimensions. Clearly the $d^2$ moduli are all contained
in
the matrix $\Xi.$
The  discrete target space duality symmetries are determined by searching for
all integer-valued linear transformations of the quantum numbers which
 leaves the spectrum invariant.
These linear transformations
can be written as
$$\Omega: u \longrightarrow S_{\Omega}(u) = \Omega^{-1} u.\eqn\lin$$
To preserve the spin, the transformation  matrix $\Omega$
should satisfy the condition:
$$\Omega^{t}\eta\Omega= \eta. \eqn\linea$$

Moreover, the invariance of $H$
induces a   transformation on the moduli.
Such a transformation defines the action of the duality group  given by
$$
\Xi \longrightarrow S_{\Omega}(\Xi) = \Omega^t \Xi \Omega.\eqn\mo$$
The generalization of the above results to the orbifold case, without Wilson
lines is
straightforward.
The orbifold is defined by the quotient of the torus by a
group of automorphisms $P$
of the lattice, also known as the point group [\one].
This group acts on the quantum numbers by
$$u \longrightarrow u^\prime = {\cal R}u, \qquad {\cal R}^N=1,\eqn\qu$$
where $\cal R$ is given by the matrix
$${\cal R}= \pmatrix{
	Q & 0 \cr
	0& {(Q^t)}^{(-1)}}\eqn\po$$
and $Q$ is an integer matrix specifying the orbifold point group.
To insure that the point group is
a lattice automorphism,
the background fields must satisfy
$${\cal R}^t\Xi {\cal R} = \Xi,\quad \Rightarrow\qquad
Q^tGQ=G,  \qquad Q^tBQ=B.\eqn\compatible$$
Finally the modular symmetries of the orbifold
are those of the torus commuting with the twist matrix ${\cal R}$ [\spalinski].

Before moving on to discuss modular symmetries of six dimensional
$Z_N$ orbifolds, it is useful to describe the target space duality groups in
2-dimensional toroidal  [\verlinde] and $Z_N $ orbifold compactifications.
In the two-dimensional case, it is convenient to group
the four real degrees of freedom parametrizing the background
into two complex moduli [\verlinde] defined as
$$T=2\Big(B+i\sqrt{\det{G}}\Big),   \qquad
U=\Big ({G_{12}\over G_{11}}+i{\sqrt{\det{G}}\over G_{11} }\Big)\eqn\complex$$
In this complex parametrization of the moduli, the duality group is given by
two copies of the modular group $PSL(2, Z)$ acting on the moduli as
$$U\rightarrow {a'U+b'\over c'U+d'}, \qquad a'd'-b'c'=1,\qquad
T\rightarrow {aT+b\over cT+d}, \qquad ad-bc=1.\eqn\hib$$
These transformations,
respectively,
are induced from  the following transformations on
the quantum numbers
$$\eqalign{\pmatrix{w\cr p}\rightarrow \Omega^{-1}_U\pmatrix{w\cr
p}=&\pmatrix{{\bf M}&0\cr 0&{({\bf M}^t)}^{-1}}\pmatrix{w\cr p},\cr
\pmatrix{w\cr p}\rightarrow\Omega^{-1}_T\pmatrix{w\cr p}=&\pmatrix{d{\bf
I}_2&-c{\bf L}\cr b{\bf L}&a{\bf I}_2} \pmatrix{w\cr p},}\eqn\transformation$$
where
$$ {\bf M}=\pmatrix{a'&-b'\cr -c'&d'},\quad {\bf L}=\pmatrix {0&1\cr
-1&0}.\eqn\trans$$
Note that the spectrum is also invariant under the exchange $T\leftrightarrow
U$ induced by the exchange $n_1\leftrightarrow -m_1$ and \lq\lq  parity
transformations \rq\rq $T\leftrightarrow -\bar T, U\leftrightarrow -\bar U$
[\verlinde].

The symmetries of the two-dimensional
${\bf Z}_N$ orbifold are those of the torus
commuting with the matrix ${\cal R}$ defining the twist. In the case when
$N=2$, one obtains
the same modular symmetries as for the toroidal case. However for $N\not =2$,
the
twist freezes the $U$ moduli and it can be easily seen that the
$PSL(2, Z)$ acting on $T$ still describes the duality group, since
$\Omega^{-1}_T$
commutes with all the ${\cal R}$ matrices defining  the  twists of the various
orbifolds.

It is now clear, from the results of the previous section,  that  in orbifold
models for which  the lattice is  a product of three two-dimensional
sublattices, and where each of the three complex planes of
the orbifold lies entirely
in each of these sublattices, the duality group is a product of $PSL(2, Z)$'s
one for each complex modulus.  Such symmetry is  demonstrated by studying the
spectrum of the untwisted sectors of these theories.  Also in these models,
the twisted sectors with one unrotated complex plane have the modular group as
a duality group. This is simply because the twisted spectrum depends on the
moduli of the unrotated plane in the same way as that of the untwisted sectors.
 Thus the threshold corrections to the gauge coupling constants in these models
are invariant under the modular group $PSL(2, Z)$ [\threshold].
However,  there are many  orbifold models where the unrotated plane does not
lie in a two-dimensional sub-lattice. Examples of such models are certain
${\bf Z}_N$ Coxeter orbifolds [\kat].
It is our purpose in this section to study the duality symmetries of the
threshold corrections in these models by investigating the symmetries of the
spectra of their  twisted sectors with one unrotated plane.

As an example, consider the orbifold
${\bf Z}_6-II$, with the twist defined by $\theta=(2,1,-3)/6$ and an
 $SU(6)\times SU(2)$ lattice.\footnote*{ the notation $(\zeta_1 , \zeta_2,
\zeta_3)$ is such
that the action of $\theta $ in the complex basis is \break
$({e}^{2 \pi {\rm i} \zeta_1},
{e}^{2 \pi {\rm i} \zeta_2}, {e}^{2 \pi {\rm i} \zeta_3}) $.}
Clearly in this model the $\theta^2$ and $\theta^3$ sector, respectively, have
the first and third planes unrotated.
We would like to investigate the symmetry  group for the moduli $T_1$ and
$(T_3, U_3)$
associated with
the first and third complex planes respectively, which leaves the spectrum of
the $\theta^2$ and $\theta^3$ twisted sectors invariant. The matrix $Q$
defining the twist action on the quantum numbers is given by
$$Q=\pmatrix{0&0&0&0&-1&0\cr
1&0&0&0&-1&0\cr 0&1&0&0&-1&0\cr 0&0&1&0&-1&0\cr 0&0&0&1&-1&0\cr 0&0&0&0&0&-1}
. \eqn\qua$$
The constant background fields compatible with the twist are obtained
using \compatible\ and are given as

$$G=\pmatrix {r^2&x&l^2&R^2&l^2&u^2\cr
	     x&r^2&x&l^2&R^2&-u^2\cr
	     l^2&x&r^2&x&l^2&u^2\cr
	     R^2&l^2&x&r^2&x&-u^2\cr
	     l^2&R^2&l^2&x&r^2&u^2\cr
	     u^2&-u^2&u^2&-u^2&u^2&y},\eqn\met$$
$$B=\pmatrix{0&-\beta&-\delta&0&\delta&-\gamma\cr
\beta&0&-\beta&-\delta&0&\gamma\cr \delta&\beta&0&-\beta&-\delta&-\gamma\cr
0&\delta&\beta&0&-\beta&\gamma\cr
-\delta&0&\delta&\beta&0&-\gamma\cr
\gamma&-\gamma&\gamma&-\gamma&\gamma&0},\eqn\anti$$
with $R^2=-2l^2-r^2-2x$.
Consider the $\theta^3$ sector first. Here the twisted states
have left and right momenta, $P_L$ and $P_R$, characterized by the winding and
momenta  $w$ and $p$ satisfying $Q^3w=w$ and ${\big ({(Q^T)}^{-1}\big)}^3p=p.$
Therefore, they are given by
$$w=\pmatrix{n_1\cr n_2\cr 0\cr n_1\cr n_2\cr 0},\qquad \qquad \qquad
p=\pmatrix{m_1\cr m_2\cr -m_1-m_2 \cr m_1\cr m_2\cr 0}.\eqn\adil$$
In order to study the duality  group of $T_1,$ it is convenient to recast the
geometry dependent scaling and spin $H_1$,  $S_1$ associated with the vertex
operators creating  the $\theta^3$  twisted states  in a quadratic form similar
to \scale.
After some algebraic calculations it turns out that one can write
$$H_1 ={1\over2}  V_1^T\pmatrix{
	 2(G_1-B_1)G_1^{-1}(G_1+B_1) & B_1G_1^{-1}\cr
		     - G_1^{-1} B_1&{1 \over 2} G_1^{-1}}V_1, \qquad S_1
		     ={1\over2}  V_1^T \eta V_1\eqn\spain$$
where
$$V_1=\pmatrix {w\cr 2p}, \  G_1=2\pmatrix{-2(l^2+x)&l^2+x\cr
l^2+x&-2(l^2+x)}, \  B_1=2\pmatrix{0&-\beta+\delta\cr
\beta-\delta&0}.\eqn\sandrine$$

Clearly the transformation
$T_1=\displaystyle{{aT_1+b\over cT_1+d}}$  leaves the theory invariant provided
that one transforms
$V_1 =\left (\matrix{w\cr p'\cr}\right ), \quad (p' = 2 p),$ as in eq.
\transformation,
$$\left (\matrix{w\cr p'\cr}\right )\rightarrow \pmatrix{
d{\bf I}_2 & -c{\bf L}\cr
b{\bf L} & a{\bf I}_2\cr}\left (\matrix{w\cr p'\cr}\right ).\eqn\eyes$$

In order that $p$ tranforms as integers, $b$ must be even. Therefore  the
modular group $\Gamma_{T_1}$ associated with the $T_1$ moduli is $\Gamma^0(2)$.
 In general the group $\Gamma^0(n)$ is represented
by the following set of matrices
$$\Gamma^0(n)=\pmatrix{a&b\cr c&d};\qquad  ad-bc=1, \quad b=0 \  \hbox{(mod)} \
 n.\eqn\todd$$
Similarly one can repeat the same analysis for the $\theta^2$ twisted sector,
here
 the twisted states
have left and right momenta, $P_L$ and $P_R$, characterized by the winding and
momenta  $w$ and $p$ satisfying $Q^2w=w$ and ${\big ({(Q^T)}^{-1}\big)}^2p=p.$
Therefore, they are given by
$$w=\pmatrix{n_1\cr 0\cr n_1\cr 0\cr n_1\cr n_2},\qquad \qquad \qquad
p=\pmatrix{m_1\cr -m_1\cr m_1 \cr -m_1\cr m_1\cr m_2}.\eqn\adam$$

The geometry dependent scaling and spins $H_3$ and spin $S_3$ associated with
the vertex operators creating  the $\theta^2$  twisted states  are given by
$$H_3={1\over2}  V_3^T\pmatrix{
	 2(G_3-B_3)G_3^{-1}(G_3+B_3) & B_3G_3^{-1}\cr
		     - G_3^{-1} B_3&{1 \over 2} G_3^{-1}}V_3, \qquad S_3
		     ={1\over2}  V_3^T \eta V_3\eqn\England$$

where
$$V_3=\pmatrix {w\cr p'}, \quad p'=\pmatrix {3&0\cr 0&1}p,
\quad G_3=\pmatrix{6l^2+3r^2&3u^2\cr 3u^2&y}, \quad
B_3=\pmatrix{0&-\gamma\cr \gamma&0}.$$
Using \complex,  $T_1$ is expressed in terms of $(G_1, B_1)$, while the moduli
$T_3$ and $U_3$ corresponding to the third plane are expressed in terms of
$(G_3, B_3)$.

Now turning to the duality symmetries for the third plane, $H_3$ and $S_3$
remain invariant under the transformations
$$U_3\rightarrow {a'U_3+b'\over c'U_3+d'}, \qquad a'd'-b'c'=1,\qquad
T_3\rightarrow {aT_3+b\over cT_3+d}, \qquad ad-bc=1, \eqn\nora$$
provided that $V_3$ transform as in eq. \transformation. Again in order for the
$p$ to transform as integers, the following constraints are obtained
$$b= 0\  \hbox{(mod)}  \  3, \qquad  c'= 0\   \hbox{( mod)}\  3. \eqn\coff$$
Thus the duality  group in this case is $\Gamma_{T_3}\times
\Gamma_{U_3}=\Gamma^0(3)\times \Gamma_0(3).$
The group $\Gamma_0(n)$ is represented
by the following set of matrices
$$\Gamma_0(n)=\pmatrix{a&b\cr c&d};\qquad  ad-bc=1, \quad c=0 \  \hbox{(mod)} \
 n.\eqn\peace$$
Now  consider the same orbifold model  but with the lattice
$SU(6)\times SU(2).$
This model has been investigated in [\stie]
with regard to the threshold corrections to the gauge coupling constants.
In this case,  the first plane, unrotated by $\theta^2$ lies, entirely in the
sub-lattice $SU(3)$  and hence the states in the $\theta^2$ twisted sector have
winding and momenta taking values  in the $SU(3)$ lattice and its dual
respectively.  Clearly, the spectrum of these states is invariant under the
full modular group acting on the $T_1$ moduli.
The third complex planes is unrotated under the $\theta^3$ action. The
$\theta^3$ twisted states have the following geometry-dependent scale and spin
which can be written in the following quadratic forms,
$$H_3 ={1\over2}  V_3^T\pmatrix{
	 2(G_3-B_3)G_3^{-1}(G_3+B_3) & B_3G_3^{-1}\cr
		     - G_3^{-1} B_3& {1 \over 2}G_3^{-1}}V_3, \qquad S_1
		     ={1\over2}  V_3^T \eta V_3\eqn\beirut$$
where
$$V_3=\pmatrix {w\cr p'}, \quad p'=\pmatrix{
1 & 1 \cr
1 & -2 \cr}\pmatrix {w\cr p}\eqn\sidon$$
Here $G_3$ and $B_3$ are the background sub-matrices defining the moduli $T_3$
and $U_3$ [\stie].

$H_3$ and $S_3$ remains invariant under the transformation
$$T_3\rightarrow {aT_3+b\over cT_3+d};\qquad ad-bc=1,$$
provided that the momenta quantum numbers transform by
$$p'\rightarrow b{\bf L}w+ap',\quad \Rightarrow p\rightarrow
{-b\over3}\pmatrix{
-1 & 2 \cr 1 & 1 \cr}w+ap \eqn\tyre$$
In order for the momenta to transform as integers, $b$ must be a multiple of
$3.$
Therefore $\Gamma_{T_3}=\Gamma^0(3).$
Also, $H_3$ and $S_3$ remains invariant under the transformation
$$U_3\rightarrow {a'U_3+b'\over c'U_3+d'};\qquad a'd'-b'c'=1,\eqn\adil$$
provided that the momenta quantum numbers transform by
$$p'\rightarrow {(M^t)}^{-1}p',\quad \Rightarrow p\rightarrow
{1\over3}\pmatrix{
2d'+2c'+b'+a' & 2d'-4c'+b'-2a' \cr d'+c'-b'-a'& d'-2c'+b'-2a'\cr}p. \eqn\lyla$$
In order for the momenta to transform as integers, the following constrains
must be imposed,
$$ (d'+c')-(a'+b')=  0\   \hbox {( mod)}\  3. \eqn\simon$$
However, in [\stie], the threshold correction for this model, when expressed in
terms of    ${U_3}'=U_3+2$,  is  invariant under $\Gamma_{{U_3}'}=\Gamma^0(3)$.
 This symmetry can be explained as follows.  By  using  \adil\ and \lyla, we
get  the following transformation
$$U_3'\rightarrow {{\cal A}U_3'+{\cal B}\over {\cal C}U_3'+{\cal D}}; \qquad
{\cal A}{
\cal D}-{\cal B}{\cal C}=1,\quad   {\cal B}= 0\   \hbox{( mod)}\  3,\eqn\lolo$$
with
$$\eqalign{{\cal A}&=d'-2c',\qquad {\cal B}=-b'-2d'+4c'+2a',\cr
{\cal C}&=-c',\qquad\qquad {\cal D}=a'+2c'.}\eqn\sharon$$

The above procedure has been applied  for all Coxeter ${\bf Z}_N$ orbifolds to
study the duality symmetries of the twisted sectors with one unrotated plane
which does not lie in a two-dimensional sub-lattice. The results are summarized
in the following table
\vskip 0.5 cm
\begintable
Orbifold|$\theta$|Lattice| Duality group\cr
$Z_4-a$| $1/4(1,1,-2)$| $SU(4)\times SU(4)$|$\Gamma_{T_3}$=$\Gamma^0(2),$
$\Gamma_{U_3}$=$PSL(2,Z)$\cr
$Z_4-b$| $1/4(1,1,-2)$|
$SU(4)\times SO(5)\times SU(2)$|$\Gamma_{T_3}$=$\Gamma^0(2),$
$\Gamma_{U_3}$=$\Gamma_0(2).$\cr
$Z_6-II -a$|$(2,1,-3)/6$|$SU(6)\times SU(2)$|$\Gamma_{T_3}$=$\Gamma^0(3)$,
$\Gamma_{U_3}=\Gamma_0(3)$, $\Gamma_{T_1}$=$\Gamma^0(2)$\cr
$Z_6-II -b$|$(2,1,-3)/6$|$SU(3)\times SO(8)$|$\Gamma_{T_3}$=$\Gamma^0(3)$,
$\Gamma_{(U_3+2)}=\Gamma^0(3)$\cr
$Z_6-II -c$|$(2,1,-3)/6$|$SU(3)\times SO(7)\times
SU(2).$|$\Gamma_{T_3}$=$\Gamma^0(3)$,$\Gamma_{U_3}=\Gamma_0(3)$\cr
$Z_8-II -a$|$(1,3,-4)/8$|$SU(2)\times SO(10)$|$\Gamma_{T_3}$=$\Gamma^0(2)$,
$\Gamma_{U_3}=\Gamma_0(2)$\cr
$Z_{12}-I-a$| $(1,-5,4)/12$|$E_6$|$\Gamma_{T_3}=\Gamma^0(2).$
\endtable
As can be seen from the various examples in the above table, the duality group
is different for
different lattice choices.
The symmetries of the threshold corrections in the examples considered in
[\stie]  are in agreement with our results.
The duality symmetries of string loop threshold corrections
of other cases in the table, also agree with the duality symmetries obtained
from
the spectra of states [\us].

In conclusion, we have calculated the symmetry groups associated with the
moduli of the $\bf Z_N$
Coxeter orbifolds  with planes that do not entirely lie in a two-dimensional
sub-lattice
of the full torus lattice. The duality groups of these models is always a
congruence subgroup of
$PSL(2,Z)$.  The form of the lattice plays an important role in the
determination of the modular
group.  It should be noted that we only considered background fields with no
Wilson lines, $i.e.$, $(2,2)$ models with unbroken $E_6$ gauge symmetry. To
make contact with the low-energy physics, one should consider $(2,0)$ models
with the  inclusion of  the appropriate  Wilson lines. These Wilson lines will
appear in the expression of the momenta of the twisted sectors with one
unrotated plane.  An important question for string phenomenology  is the
determination
of the target space duality symmetry  in the presence of Wilson lines and the
calculation of the threshold corrections for  $(2,0)$ orbifolds.
\vskip 1cm
\centerline{ ACKNOWLEDGEMENTS}
This work was supported in part by the S.E.R.C. and the work of  S. T. was
supported by the Royal Society.
\vfill\eject

 \vfill\eject
\refout
\end

\end